\documentclass[11pt]{revtex4}
\usepackage{graphicx}
\usepackage{amsmath}
\usepackage{amssymb}

\input{tcilatex}
\begin{document}

\title{Parametric Oscillation of a Moving Mirror Driven by Radiation
Pressure in a Superconducting Fabry-Perot Resonator System}
\author{{\large Raymond Y. Chiao$^a$, Luis A. Martinez$^b$, Stephen J.
Minter$^c$, and Alexey Trubarov$^c$} \\ \vspace{0.1in} {\scriptsize {$^a$University of California, Merced,
Schools of Natural Sciences and Engineering, P.O. Box 2039, Merced, CA
95344, USA \\
Corresponding author, email: rchiao@ucmerced.edu \\
\vspace{0.05in} $^b$University of California, Merced, School of Natural
Sciences, P.O. Box 2039, Merced, CA 95344, USA \\
\vspace{0.05in} $^c$Vienna Center for Quantum Science and Technology,
Faculty of Physics, \\
University of Vienna, Boltzmanngasse 5, A-1090 Vienna, Austria\\}}
\vspace{0.1in}\normalsize{PACS: 42.50.Pq, 42.79.Gn, 07.57.Hm, 03.70.+k, 11.10.-z, 03.65.-w, 42.50.Xa}}

\begin{abstract}\textbf{Abstract}: A moving pellicle superconducting mirror, which is driven by radiation pressure on its one side, and by the Coulomb force on its other side, can become a parametric oscillator that can generate microwaves when placed within a high-Q superconducting Fabry-Perot resonator system. A paraxial-wave analysis shows that the fundamental resonator eigenmode needed for parametric oscillation is the TM$_{011}$ mode. A double Fabry-Perot structure is introduced to resonate the pump and the idler modes, but to reject the parasitic anti-Stokes mode. The threshold for oscillation is estimated based on the radiation-pressure coupling of the pump to the signal and idler modes, and indicates that the experiment is feasible to perform. 
\end{abstract}

\maketitle

\address{\vspace{0.1in} {\scriptsize {$^a$University of California, Merced,
Schools of Natural Sciences and Engineering, P.O. Box 2039, Merced, CA
95344, USA }\\
Corresponding author, email: rchiao@ucmerced.edu \\
\vspace{0.05in} $^b$University of California, Merced, School of Natural
Sciences, P.O. Box 2039, Merced, CA 95344, USA \\
\vspace{0.05in} $^c$Vienna Center for Quantum Science and Technology,
Faculty of Physics, \\
University of Vienna, Boltzmanngasse 5, A-1090 Vienna, Austria\\
}}

\section{Introduction}

\begin{figure}[tbh]
\centering
\includegraphics[angle=0,width=.4\textwidth]{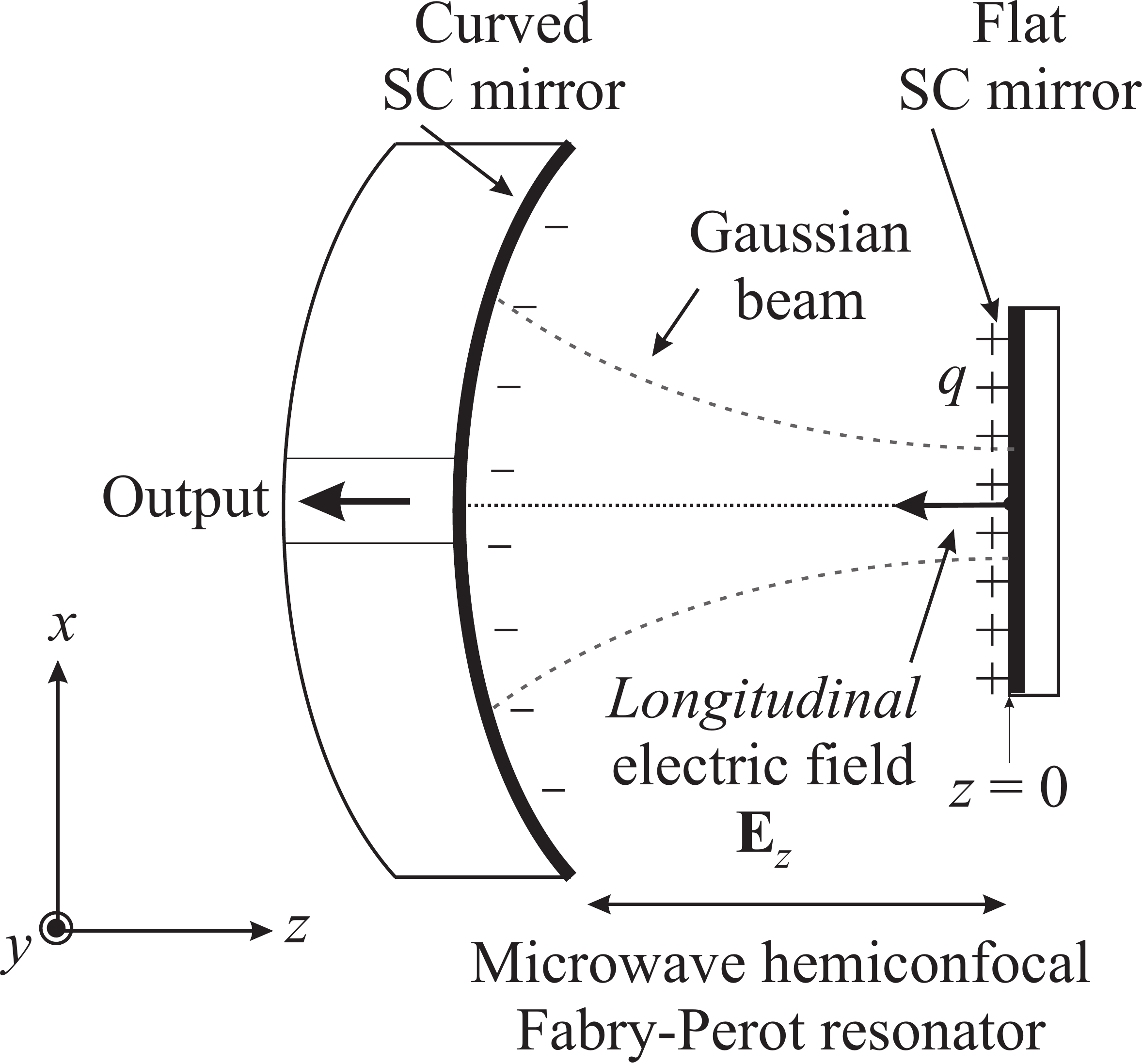}
\caption{Hemiconfocal Fabry-Perot superconducting (SC) resonator with
electrostatically charged mirrors, for generating microwaves in a parametric
oscillator. A Gaussian beam mode focused by a curved mirror onto a flat
mirror with charge $q$ that can move back and forth parallel to the $z$
axis, yields a longitudinal electric field $\mathbf{E}_z$ satisfying
conducting boundary conditions. The lowest mode of the resonator is the
transverse magnetic TM$_{011}$ mode (see Appendix A).}
\label{Longitudinal-component-of-E-field-in-TM-mode}
\end{figure}

Parametric oscillators for generating electromagnetic microwaves might be
possible, based upon the idea that a moving mirror is like a moving piston
that can perform work \emph{nonadiabatically} on radiation contained within
a cavity. Above a certain threshold, the parametric action of the moving
mirror will exponentially amplify this radiation until it can become a
large-amplitude, classical wave. An essential component for such a
parametric oscillator is a high-Q superconducting (SC) resonator, which can
be formed by a fixed, spherical curved SC mirror in conjunction with a
moving, flat SC\ mirror (see Figure 1).

To understand how the proposed parametric oscillator works, let us imagine
performing a thought experiment in which some weak, on-resonance,
\textquotedblleft seed\textquotedblright\ microwave radiation from an
outside microwave source, is injected backwards through the output hole of
the curved SC mirror of Figure 1, into the volume of the cavity between the
curved and flat mirrors. Thus the SC resonator can be filled with
\textquotedblleft seed\textquotedblright\ radiation, exciting it into its
fundamental TM$_{011}$ eigenmode. As a result, there will appear a
longitudinal electric field $\mathbf{E}_{z}$ at the surface of the SC flat
mirror that will be oscillating at the TM$_{011}$ eigenmode frequency, i.e.,
at a microwave frequency (see Appendix A).

Next, imagine that the flat SC mirror consists of a thin SC film sputtered
onto the left side of a thin, light, flexible diaphragm (e.g., the pellicle
mirror to be introduced later in Figure 4), which is sufficiently thin so
that this diaphragm can easily be driven into mechanical motion. Moreover,
imagine that this film is electrostatically charged with a DC charge $q$.
Then the longitudinal electric field $\mathbf{E}_{z}$ at the surface of the
SC film will lead to a force%
\begin{equation}
\mathbf{F}_{z}\left( t\right) =q\mathbf{E}_{z}\left( t\right)
\label{F_z  prop to E_z}
\end{equation}%
oscillating at the $same$ microwave frequency as that of $\mathbf{E}%
_{z}\left( t\right) $. Note that (\ref{F_z prop to E_z}) is a \emph{linear}
relationship between the force $\mathbf{F}_{z}\left( t\right) $ and the
electric field $\mathbf{E}_{z}\left( t\right) $. In this way, a force
oscillating at a microwave frequency can be exerted upon the diaphragm
parallel to the $z$ axis, and the flat SC mirror will be driven into simple
harmonic motion at the same microwave frequency as that of the TM$_{011}$
eigenmode of the SC resonator of Figure 1.

Now consider the opto-mechanical configuration sketched in Figure 2, in
which a laser beam from the right is incident on a moving mirror (e.g., on
the flat SC mirror of Figure 1 with multilayer dielectric optical coatings
deposited on its right side). When this moving optical mirror is combined
with a fixed optical mirror in an optical Fabry-Perot-cavity configuration,
there will arise a production of Doppler sidebands, which can then be
utilized either for the laser cooling of the moving mirror by means of a 
\emph{red-detuned} laser tuned to the lower Doppler sideband \cite%
{Aspelmeyer JOSA B}, or for the parametric oscillation of the moving mirror
excited in an elastic mode at acoustical frequencies by means of a \emph{%
blue-detuned} laser tuned to the upper Doppler sideband \cite{MIT paramp}.
Above the threshold for parametric oscillation of the moving mirror within
the SC resonator of Figure 1, a \textquotedblleft signal\textquotedblright\
wave would begin to build up, growing exponentially with time starting from
the injected \textquotedblleft seed\textquotedblright\ microwave radiation.
However, once parametric oscillation above threshold has occurred, one could
turn off the source of the \textquotedblleft seed\textquotedblright\
radiation. The SC resonator would then continue to oscillate as an
autonomous source of the same microwaves as the \textquotedblleft
seed\textquotedblright\ in the absence of the \textquotedblleft
seed\textquotedblright , just like the autonomous generation of microwave
radiation by the original ammonia maser above its oscillation threshold \cite%
{GZT}.

\begin{figure}[tbh]
\centering
\includegraphics[angle=0,width=.4\textwidth]{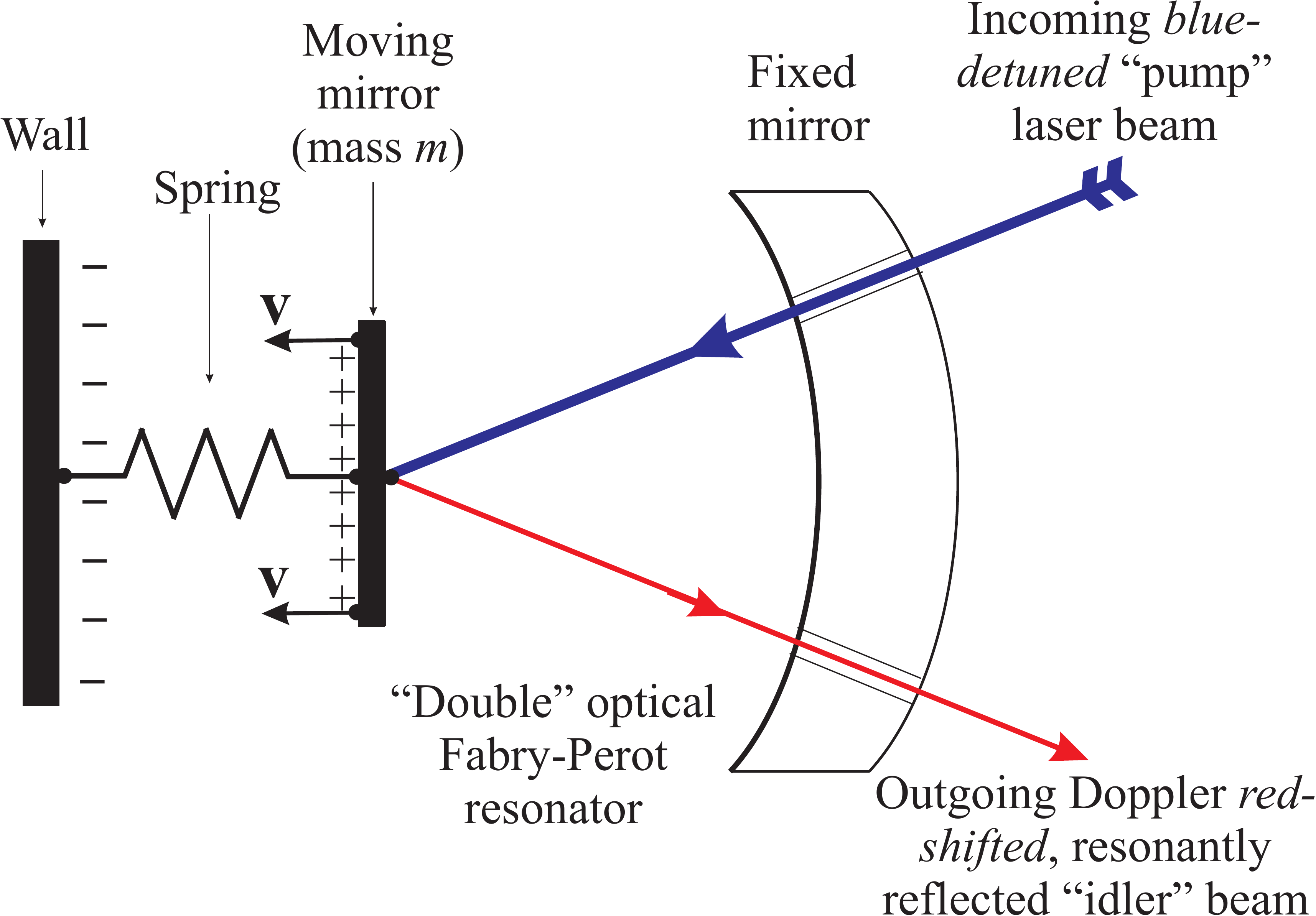}
\caption{A mass-and-spring model for an opto-mechanical parametric
oscillator that could generate microwave radiation starting from
\textquotedblleft seed\textquotedblright\ radiation. The SC Fabry-Perot
microwave resonator in Figure 1 is modeled by a charged mass attached via a
spring to a fixed, charged wall. The \textquotedblleft
pump\textquotedblright\ laser beam (in blue) is \emph{blue-detuned} so as to
coincide with the \emph{upper} Doppler sideband arising from the simple
harmonic motion of the moving mirror. After reflection from the moving
mirror, the laser beam is red-shifted to become the \textquotedblleft
idler\textquotedblright\ beam (in red), whose frequency coincides with the 
\emph{lower} Doppler sideband. The \textquotedblleft pump\textquotedblright\
laser beam will exert a radiation pressure force that will drive the mirror
into parametric oscillations. Here the \textquotedblleft
Double\textquotedblright\ optical Fabry-Perot resonator represents a
simplified model for the doubly-resonant structure depicted in Figure 3,
which serves to resonate both the \textquotedblleft idler\textquotedblright\
and the \textquotedblleft pump\textquotedblright\ frequencies, but to reject
the any Doppler up-shifted frequencies.}
\label{fig:Optomechanical-laser-heating-(flipped)}
\end{figure}

\section{Opto-mechanical Model for Parametric Oscillators}

For ease of understanding, consider the simplified opto-mechanical model for
the parametric process shown in Figure 2, in which a moving mirror attached
to a spring is coupled via radiation pressure to a strong \textquotedblleft
pump\textquotedblright\ laser at frequency $\omega _{p}$, a weak
\textquotedblleft idler\textquotedblright\ light wave within an optical
resonator at frequency $\omega _{i}$, and a weak \textquotedblleft
signal\textquotedblright\ wave in the SC resonator at frequency $\omega _{s}$%
, which is being represented as a simple harmonic oscillator with a
resonance frequency $\omega _{s}$, such that%
\begin{equation}
\omega _{p}=\omega _{i}+\omega _{s}
\end{equation}%
The energy for the oscillating fields at $\omega _{i}$ and $\omega _{s}$
comes from the energy supplied by the pump laser at $\omega _{p}$.

In Figure 2, we have replaced the hemiconfocal microwave resonator shown in
Figure 1 by a mass-and-spring model, in which the simple harmonic mechanical
motion of the flat SC mirror of Figure 1 driven in the presence of the
charge $q$ by the longitudinal electric field $\mathbf{E}_{z}\left( t\right) 
$ parallel to the $z$ axis, is modeled by the simple harmonic motion of a
mass attached via a spring to the fixed wall on the left side of Figure 2.
The justification for using this mechanical model (see Appendix B) is that
we have found that the fundamental eigenmode solution of the SC hemiconfocal
microwave resonator, which is the Gaussian-beam TM$_{011}$ mode, possesses a
longitudinal component $\mathbf{E}_{z}$ of the electric field which has a
nonvanishing component along the $z$ axis of the resonator acting on the
electrostatic charge $q$\ at the surface of the flat mirror of Figure 1.
Therefore there exists a longitudinal component of the force that drives the
flat mirror back and forth into simple harmonic motion along the $z$ axis,
whose motion can then be simulated by the mass-and-spring model of Figure 2.

\begin{figure}[tbh]
\centering
\includegraphics[angle=0,width=.4\textwidth]{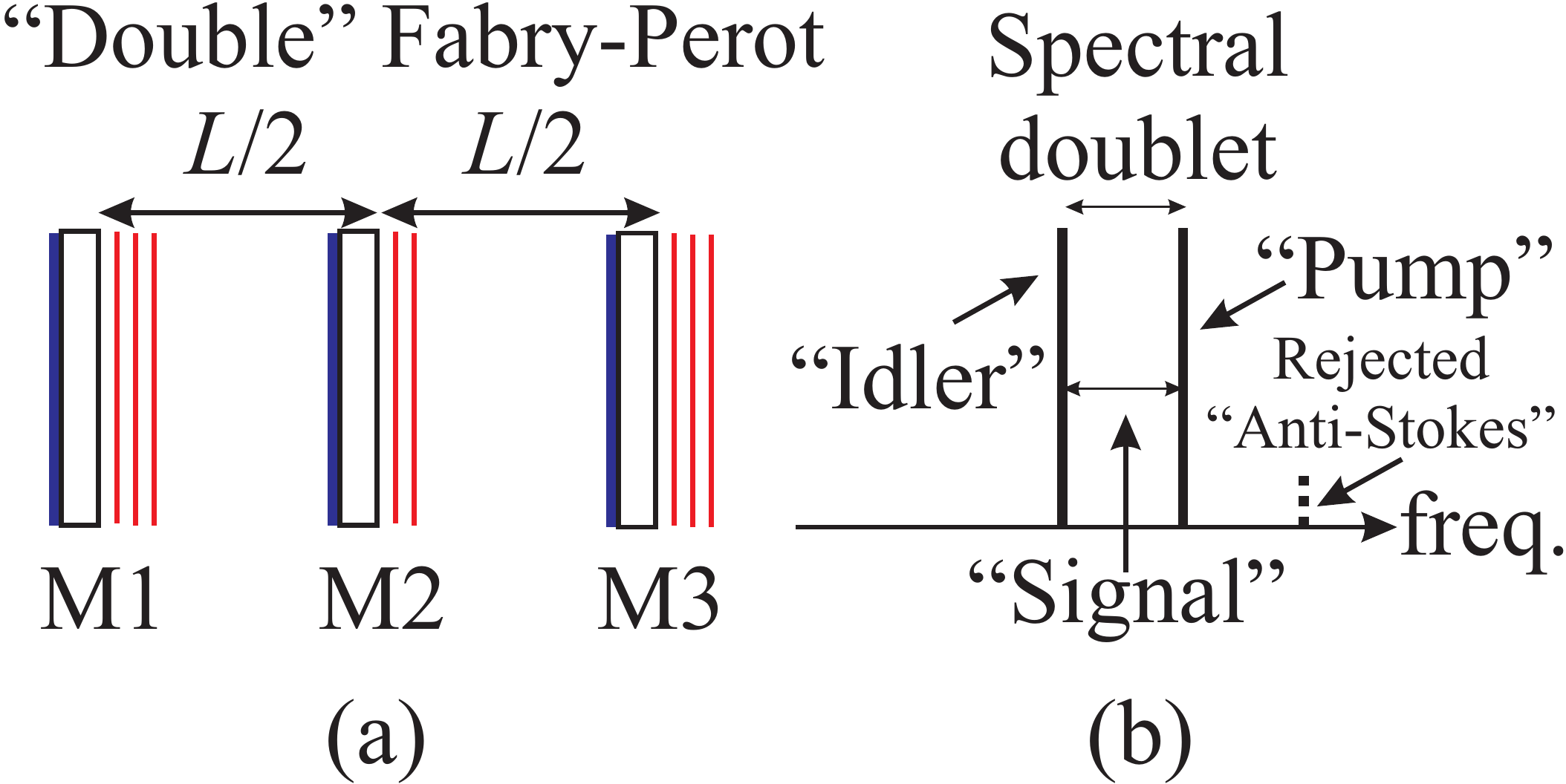}
\caption{(a) \textquotedblleft Double\textquotedblright\ Fabry-Perot is a
three-mirror resonator system with three equally spaced mirrors M1, M2, and
M3. The leftmost mirror M1 is the flat, moving mirror attached to the spring
of Figure 2, i.e., the \textquotedblleft pellicle-mirror\textquotedblright\
of Figure 4. The pump laser is incident from the right through the rightmost
mirror M3. Multi-layer dielectric coatings are indicated by red lines;
anti-reflection coatings by blue lines. The reflectivity of M2 is less than
that of M1 and M3. (b) The spectral transmission of this resonator system
has a doublet structure, whose splitting is proportional to the
transmittivity of the middle mirror M2. The frequency of the
\textquotedblleft pump\textquotedblright\ laser of the parametric oscillator
coincides with the upper member of the doublet, and the frequency of the
\textquotedblleft idler\textquotedblright\ (\textquotedblleft
Stokes\textquotedblright) wave coincides with the lower member. The doublet
splitting frequency is resonant with the \textquotedblleft
signal\textquotedblright\ frequency of the parametric oscillator. The
parasitic \textquotedblleft Anti-Stokes\textquotedblright\ wave is rejected
by the structure.}
\label{fig:Double-Fabry-Perot}
\end{figure}

Furthermore, we shall model, by means of a single, simplified Fabry-Perot
structure on the right side of Figure 2, a \textquotedblleft
Double\textquotedblright\ optical Fabry-Perot resonator, which is depicted
in more detail in Figure 3, whose purpose is to simultaneously resonate both
the strong, incoming \emph{blue-detuned} laser (i.e., the \textquotedblleft
pump\textquotedblright\ laser) and the weak, Doppler red-shifted
\textquotedblleft idler\textquotedblright\ light wave produced upon
reflection from the moving mirror, but whose purpose is also to serve as a
rejection filter to reject any undesirable Doppler blue-shifted (or
\textquotedblleft anti-Stokes\textquotedblright ) light. Later, we shall put
all these pieces of the parametric oscillator system together in Figure 4.

\section{The \textquotedblleft Double\textquotedblright\ Fabry-Perot
Resonator}

Figure 3(a) shows the detailed structure of the \textquotedblleft
Double\textquotedblright\ Fabry-Perot resonator, which consists of three
equally spaced mirrors M1, M2, and M3. Suppose that M1 is the moving mirror
of Figure 2 (or the \textquotedblleft pellicle mirror\textquotedblright\ of
Figure 4). Suppose further that at $t=0$ light initially fills the left half
of the structure between M1 and M2, but that there is initially no light in
the right half between M2 and M3. Then after a period of time determined by
the transmittivity of the middle mirror M2, light will leak from the left
cavity to the right cavity, until light fills the right cavity. This results
in a periodic sloshing back and forth of the light between the left and
right halves of the structure at a frequency equal to the splitting
frequency of the spectral doublet shown in Figure 3(b). The
\textquotedblleft pump\textquotedblright\ laser will be tuned into the
resonance with the upper member of the spectral doublet; the
\textquotedblleft idler\textquotedblright\ wave to the lower member of this
doublet. Any small amount of noise at the idler frequency will lead to a
radiation pressure force on the moving mirror M1 that will be modulated at
the doublet splitting frequency, which can be chosen to be tuned into
resonance with the SC microwave resonator frequency of Figure 1. This then
can lead to a mutual, resonant reinforcement of the noise in the
\textquotedblleft idler\textquotedblright\ and \textquotedblleft
signal\textquotedblright\ modes that leads to exponential growth of both by
parametric amplification, and possibly to oscillation above a certain
threshold, as will be presently shown.

An important feature of the \textquotedblleft Double\textquotedblright\
Fabry-Perot is that it automatically rejects all Doppler up-shifted, or
\textquotedblleft anti-Stokes,\textquotedblright\ spectral components
arising from the motion of the moving mirror towards the laser in Figure 2.
Such undesirable \textquotedblleft anti-Stokes,\textquotedblright\ or
up-shifted, frequency components of the light can rob energy away from the
desired \textquotedblleft Stokes,\textquotedblright\ down-shifted, frequency
components, specifically, the \textquotedblleft idler\textquotedblright\
frequency necessary for exponential parametric gain to occur. Since the
spectral doublet structure of Figure 3(b) does not have any resonance at the
\textquotedblleft anti-Stokes\textquotedblright\ frequency, the
\textquotedblleft Double\textquotedblright\ Fabry-Perot will serve as a
spectral rejection filter that prevents light from building up inside the
resonator at any unwanted, Doppler up-shifted frequencies.

\section{Pellicle-Mirror Fabry-Perot System}

\begin{figure}[tbh]
\centering
\includegraphics[angle=0,width=.4\textwidth]{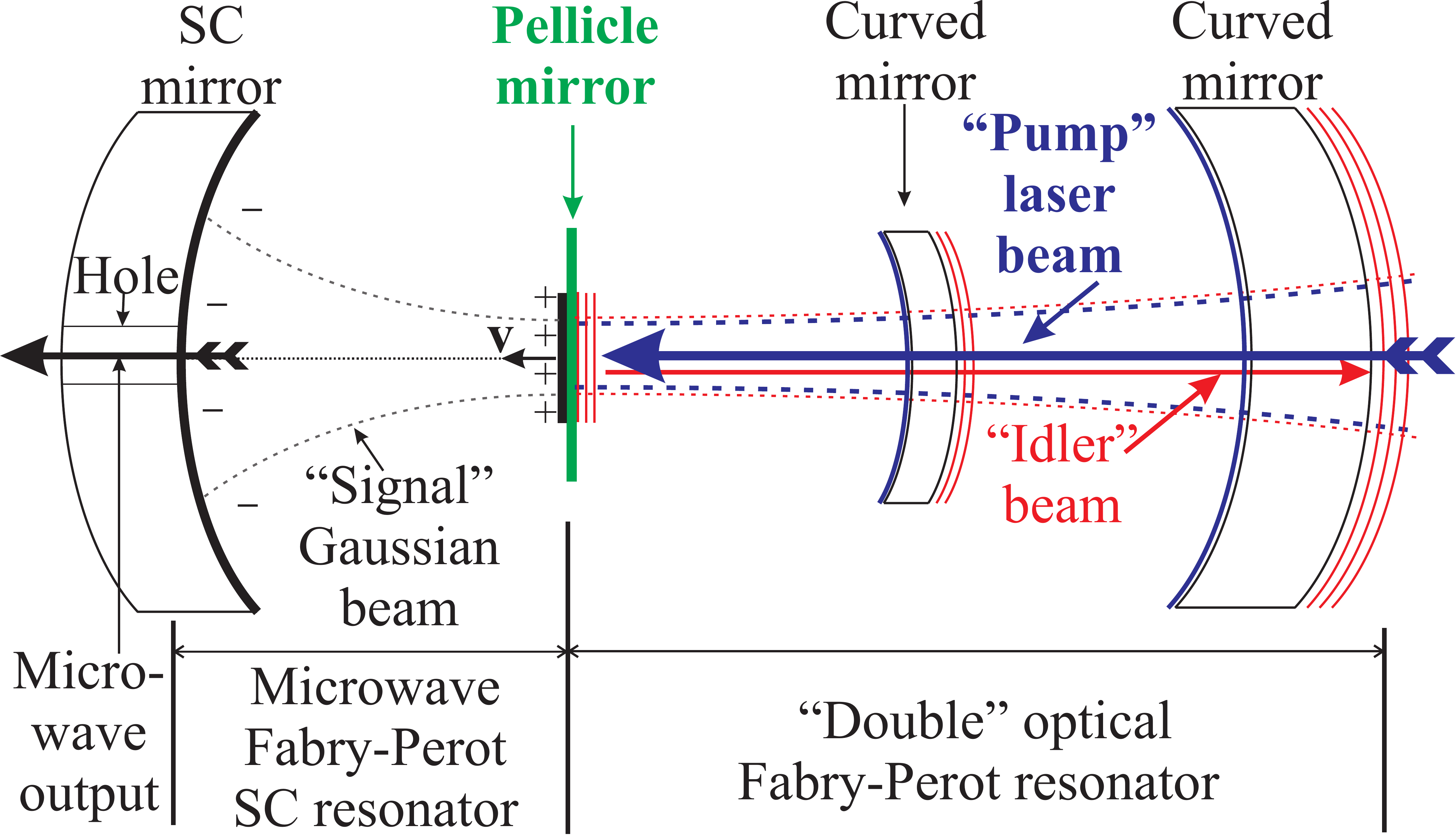}
\caption{Pellicle-mirror parametric oscillator for generating microwaves. A
thin, charged SC niobium film (in black) is sputtered onto the left side of
a thin pellicle substrate (in green). A thin, oppositely charged SC
niobium film (also in black) is sputtered onto the reflecting surface of the
leftmost curved SC mirror, in order to form an extremely high-Q SC
resonator. Optical coatings (in red) are evaporated onto the right side of
the pellicle mirror. Thus the pellicle mirror divides the entire structure
into two halves. In the left half, the microwave-frequency Fabry-Perot
resonator is for resonating a microwave \textquotedblleft
signal\textquotedblright\ within a Gaussian beam (in dotted black). In the
right half, the \textquotedblleft Double\textquotedblright\ optical
Fabry-Perot resonator is for resonating both the \textquotedblleft
idler\textquotedblright\ light beam (in dashed red) and the blue-detuned
\textquotedblleft pump\textquotedblright\ laser beam (in dashed blue). The
microwave mode on the left side of the pellicle is the TM$_{011}$ mode, and
the optical mode on the right side of the pellicle is the TEM$_{00n}$ mode.}
\label{fig:Triple-FP-with-pellicle-mirror}
\end{figure}

A practical implementation of the \textquotedblleft moving
mirror\textquotedblright\ idea at the heart of the microwave parametric
oscillator, is to evaporate a thin SC film onto a thin, flexible, low-mass
pellicle (with a thickness of around two microns \cite{Maev}) stretched
tautly over a circular wire frame in order to form a \textquotedblleft
drumhead\textquotedblright\ with a low-frequency acoustical eigenmode whose
mode pattern has a central maximum at the center of the pellicle. On the
reverse side of the pellicle, one could then evaporate dielectric optical
coatings to reflect both the strong \textquotedblleft
pump\textquotedblright\ laser light beam, and the weak \textquotedblleft
idler\textquotedblright\ light beam, for implementing the \textquotedblleft
Double\textquotedblright\ Fabry-Perot cavity for the \textquotedblleft
pump\textquotedblright\ and \textquotedblleft idler\textquotedblright\
optical\ beams, which is sketched in Figure 3.

In Figure 4 (which is not drawn to scale, since the diffraction of the
microwave-frequency \textquotedblleft signal\textquotedblright\ Gaussian
beam will occur much more quickly than the diffraction of the
optical-frequency pump and idler light beams), we put together all the
pieces of the parametric oscillator system based on the \textquotedblleft
pellicle mirror\textquotedblright\ idea. Since microwave frequencies are
very much higher than the acoustical frequencies of the pellicle drumhead,
the motion of the central portion of the pellicle is essentially like that
of a free body (i.e., a free mass) being driven at microwave
frequencies.

One can understand the parametric amplification arising in this
configuration as follows: When the \textquotedblleft pump\textquotedblright\
laser beam enters from the right into the \textquotedblleft
Double\textquotedblright\ optical Fabry-Perot system, the pellicle will act
as if it were a piston that is being driven by the radiation pressure force
which is varying at the beat note frequency between the strong
\textquotedblleft pump\textquotedblright\ laser beam (in blue), whose
frequency is $\omega _{p}$, and the weak \textquotedblleft
idler\textquotedblright\ (or \textquotedblleft Stokes\textquotedblright )
beam (in red), whose frequency is $\omega _{i}$, such that this beat note
frequency equals the \textquotedblleft signal\textquotedblright\ frequency $%
\omega _{s}$, which is the TM$_{011}$ eigenmode resonance frequency of the
left-hand-side microwave Fabry-Perot resonator. Then during the parametric
amplification process, one quantum of the pump wave will break up into one
quantum of the signal wave plus one quantum of the idler of wave, satisfying
the energy conservation relationship%
\begin{equation}
\hbar \omega _{p}=\hbar \omega _{i}+\hbar \omega _{s}
\end{equation}%
where $\hbar \omega _{p}$ is one quantum of energy of the \textquotedblleft
pump\textquotedblright\ mode, $\hbar \omega _{i}$ is one quantum of energy
of the \textquotedblleft idler\textquotedblright\ mode, and $\hbar \omega
_{s}$ is one quantum of energy of the \textquotedblleft
signal\textquotedblright\ mode.

The moving SC pellicle mirror, when it is viewed as if it were a moving
piston, will do work on the \textquotedblleft seed\textquotedblright\
microwaves contained inside the left Fabry-Perot as it moves
nonadiabatically \cite{nonadiabatic}. Hence the action of the moving
pellicle will amplify this \textquotedblleft seed\textquotedblright\
radiation, which will in turn amplify the motion of the pellicle when it is
viewed as a moving optical mirror, because this motion will further amplify
the strength of the \textquotedblleft Stokes\textquotedblright\ Doppler
sideband of the incoming \textquotedblleft pump\textquotedblright\ laser
beam, thus amplifying the strength of the \textquotedblleft
idler\textquotedblright\ beam. This in turn increases the strength of the
beat note in the radiation pressure force acting on the pellicle, which
further increases the amplitude of the motion of the pellicle, etc., in a
feedback process. In this way, there will be a mutual reinforcement of the
\textquotedblleft signal\textquotedblright\ wave and the \textquotedblleft
idler\textquotedblright\ wave, so that an exponential growth of both waves
(above a certain threshold of oscillation) will result, as in a traditional
laser. For each quantum produced in the \textquotedblleft
idler\textquotedblright\ mode, a quantum of the \textquotedblleft
signal\textquotedblright\ mode will be produced as well, in agreement with
the Manley-Rowe relations \cite{ManleyRowe}.

\section{Threshold for Parametric Amplification}

We make a simple argument to estimate the threshold for parametric
amplification of the pellicle mirror parametric oscillator shown in Figure
4. Let us assume that the microwave cavity on the left side of the
parametric amplifier can be modeled as a mirror with a spring attached to a
fixed wall (see Figure 2) so that it forms a simple harmonic oscillator with
resonant frequency $\omega _{s}$, effective mass $m$, and quality factor $%
Q_{s}$. The right half of the parametric oscillator is regarded as a
\textquotedblleft Double\textquotedblright\ optical Fabry-Perot resonator
with two resonances at $\omega _{i}$ and $\omega _{p}$. The transmission of
this \textquotedblleft Double\textquotedblright\ Fabry-Perot has been
studied in \cite{vandestadt85,hogeveen86} and is illustrated in Figure 5.
Note that the splitting between the double peaks depends on the values of
the reflection coefficients. We require that $r_{1}=r_{3}\equiv r$ and $%
r_{2}<r$, where $r$ corresponds to reflection coefficient of the two end
mirrors, and $r_{2}$ to the middle mirror in Figure 3(a). Furthermore, the
purpose of the \textquotedblleft Double\textquotedblright\ Fabry-Perot is to
allow for the selection of the two desired optical modes $\omega _{i}$ and $%
\omega _{p}$, as already discussed. Since the \textquotedblleft
Double\textquotedblright\ Fabry-Perot acts as a single Fabry-Perot with two
closely spaced resonances, we treat the parametric amplifier as a single
Fabry-Perot cavity with a harmonically moving end mirror as illustrated in
Figure 2.

\begin{figure}[tbh]
\centering
\includegraphics[angle=0,width=.4\textwidth]{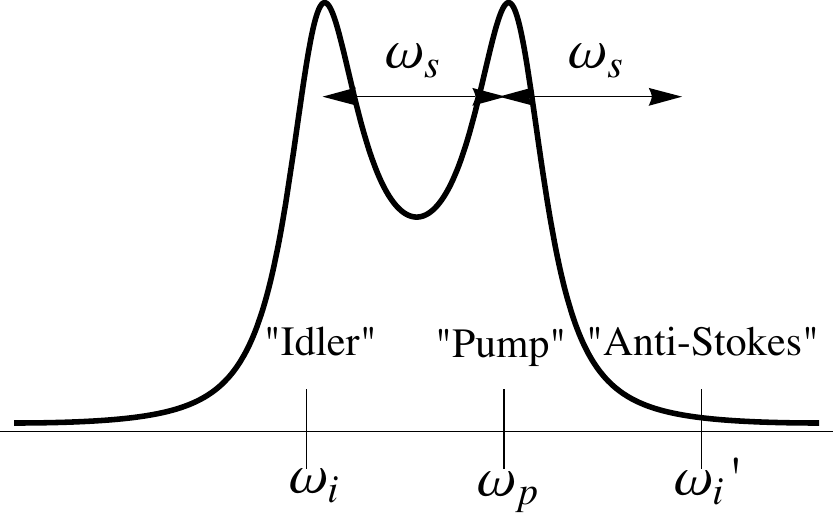}
\caption{The transmission of a \textquotedblleft Double\textquotedblright\
Fabry-Perot has resonances at the \textquotedblleft idler\textquotedblright\
mode $\protect\omega_i$ and the \textquotedblleft pump\textquotedblright\
mode $\protect\omega_p$. The difference frequency between \textquotedblleft
pump\textquotedblright\ and \textquotedblleft idler\textquotedblright\ is
resonant with the \textquotedblleft signal\textquotedblright\ frequency\ $%
\left(\rm{i.e.},\ \protect\omega_p-\protect\omega_i=\protect\omega%
_s\right)$. The \textquotedblleft anti-Stokes\textquotedblright\ $\protect%
\omega_i^{\prime }$ is off resonance and, hence suppressed.}
\label{fig:Dspectrum}
\end{figure}


The Fabry-Perot with a harmonically moving end-mirror (see Appendix C)
yields the following equations of motion 
\begin{align}
& \frac{dX}{dt}+\gamma _{s}X=\frac{i\epsilon _{0}\mathcal{A}}{2\omega _{s}m}%
\mathcal{E}_{0i}^{\ast }\mathcal{E}_{0p}e^{-i\Delta \omega t}  \label{eq:meomd} \\
& \frac{d\mathcal{E}_{0i}}{dt}+\gamma _{i}\mathcal{E}_{0i}=\frac{i}{\tau }%
X^{\ast }\mathcal{E}_{0p}e^{-i\Delta \omega t}
\end{align}%
where $\Delta \omega \equiv \omega _{p}-\omega _{i}-\omega _{s}$, $\mathcal{A%
}$ is the cross sectional area of the resonator, $X$ is a slowly varying
oscillator amplitude, $\tau $ is the round trip time between the two mirrors
of Figure 2, $\mathcal{E}_{0i}$ is the slowly varying amplitude of the
\textquotedblleft idler\textquotedblright\ mode, and $\gamma _{i}$, $\gamma
_{s}$ are the HWHM for the \textquotedblleft idler\textquotedblright\ and
\textquotedblleft signal\textquotedblright , respectively. The problem of a
Fabry-Perot with a harmonically moving end-mirror has been solved in detail
by \cite{braginsky01}. Choosing the same normalization (see Appendix C) as
in \cite{braginsky01}, we arrive at the same results 
\begin{align}
& \frac{dX}{dt}+\gamma _{s}X=\frac{i\omega _{p}\omega _{i}}{m\omega _{s}L}D_{p}D_{i}^{\ast }e^{-i\Delta \omega t} \\
& \frac{dD_{i}}{dt}+\gamma _{i}D_{i}=\frac{i\omega _{p}}{L}%
X^{\ast }D_{p}e^{-i\Delta \omega t}  \label{eq:beomd}
\end{align}%
where $D_{i}\propto \mathcal{E}_{0i}$ and $D_{p}\propto \mathcal{E}_{0p}$, and $L$ is the length
of the Fabry-Perot resonator of Figure 2. It was shown in \cite{braginsky01}
that parametric amplification at resonance occurs if 
\begin{equation}
\frac{2U_{p}Q_{i}Q_{s}}{mL^{2}\omega _{s}^{2}}>1
\end{equation}%
where $U_{p}$ is the energy stored in the \textquotedblleft
pump\textquotedblright\ mode, and the $Q_{i}$ and $Q_{s}$ are the quality
factors of the \textquotedblleft idler\textquotedblright\ and
\textquotedblleft signal\textquotedblright\ modes, respectively. We can
express this in terms of the outside laser power of the \textquotedblleft
pump\textquotedblright\ beam ($\mathcal{P}_{\text{outside}}$) by noting the
circulating power inside ($\mathcal{P}_{\text{inside}}$) the Fabry-Perot
cavity for the \textquotedblleft pump\textquotedblright\ mode is to a good
approximation 
\begin{equation}
\mathcal{P}_{\text{inside}}=\frac{\mathcal{F}}{\pi }\mathcal{P}_{\text{%
outside}}
\end{equation}%
where $\mathcal{F}$ is the finesse of the Fabry-Perot resonator, assuming a
high reflection coefficient for its mirrors. Solving for the outside
\textquotedblleft pump\textquotedblright\ intensity gives the threshold
condition for parametric amplification 
\begin{equation}
\mathcal{P}_{\text{threshold}}=\frac{\pi c}{2\mathcal{F}}\frac{mL\omega
_{s}^{2}}{Q_{i}Q_{s}}
\end{equation}%
where $L$ is the length of the Fabry-Perot, and $c$ is the speed of light.
For estimates of the threshold we select a reflection coefficient \cite%
{luisnote0} $r=0.9999$ for the two end mirrors, and $r_{2}=0.5$ for the
middle mirror (M2 of Figure 3). These values yield a spacing between the two
peaks in Figure 5 of $f_{s}\approx 10$ GHz, which also corresponds to a
finesse of $\mathcal{F}\approx 15700$, a $Q_{i}\approx 4.5\times 10^{8}$ for
a 700 nm wavelength; the quality factor of the \textquotedblleft
signal\textquotedblright\ is assumed \cite{luisnote1,kuhr07,RFbook} to be $%
Q_{s}\sim 10^{10}$, $L=1$ cm, and $m$ is assumed to be a free mass \cite%
{luisnote2} on the order of $m=2$ mg. With these parameters we find a laser
power threshold of 
\begin{equation}
\mathcal{P}_{\text{threshold}}=530\text{ mW}
\end{equation}%
It is strongly noted that these estimates are based on the ideal model
described above. However, it suffices as an order of magnitude estimate and
shows that parametric amplification in such a scheme may be possible.
Furthermore, parametric amplification of acoustic modes has already been
experimentally observed within Fabry-Perot cavities \cite%
{corbitt06,kippenberg05}.

\section{The Case of an All-microwave, All-superconducting Resonator System}

Now suppose that one were to replace the optical \textquotedblleft
Double\textquotedblright\ Fabry-Perot system in Figure 4 by a microwave
\textquotedblleft Double\textquotedblright\ all-superconducting Fabry-Perot
system, so that the entire system becomes an all-microwave,
all-superconducting resonator system. Let us also replace the incoming laser
\textquotedblleft pump\textquotedblright\ beam by an incoming microwave
\textquotedblleft pump\textquotedblright\ beam coming in from the right, and
let us then recalculate the pump threshold power needed for parametric
oscillation. For the case of microwave resonators having lengths comparable
to the microwave wavelength, the cavity finesse $\mathcal{F}$ becomes
comparable to the quality factor $Q_{p}$ for the \textquotedblleft
pump\textquotedblright\ resonator, so that now the estimate for the
threshold power becomes%
\begin{equation}
\left. \mathcal{P}_{\text{threshold}}\right\vert _{\text{pump}}^{\text{%
microwave}}\simeq c\frac{mL\omega _{s}^{2}}{Q_{p}Q_{i}Q_{s}}
\label{microwave threshold power}
\end{equation}%
Since the system is now an all-superconducting system, we shall assume that 
\cite{luisnote1,kuhr07,RFbook}%
\begin{equation}
Q_{p}\approx Q_{i}\approx Q_{s}\approx 10^{10}
\end{equation}%
Again let us assume a mass $m$ of the pellicle mirror to be 2 mg, and a
length $L$ of the resonator to be 1 cm, which corresponds to a
\textquotedblleft signal\textquotedblright\ frequency $\omega _{s}$\ of $%
2\pi \times 15$ GHz, and a \textquotedblleft pump\textquotedblright\
frequency $\omega _{p}$ of $2\pi \times 30$ GHz, in the case of a degenerate
parametric oscillator for which $\omega _{s}=\omega _{i}=\omega _{p}/2$. One
concludes from (\ref{microwave threshold power}) that%
\begin{equation}
\left. \mathcal{P}_{\text{threshold}}\right\vert _{\text{pump}}^{\text{%
microwave}}\simeq 0.05\text{ }\mu \text{W}
\end{equation}%
Therefore only microwatt-scale \textquotedblleft pump\textquotedblright\
threshold powers are needed for parametric oscillation for an all-microwave,
all-superconducting resonator system. This is clearly a feasible experiment
to perform.

\section{Appendix A: Gaussian Mode of a SC Hemiconfocal Resonator with
Longitudinal Electric Fields}

Here we examine in detail the properties of the fundamental Gaussian-beam
mode of the electromagnetic hemiconfocal resonator sketched in Figure 1,
following the methodology of the analysis of Ince-Gaussian beams introduced
in reference \cite{Ince-Gaussian}.

In order to satisfy the conducting boundary conditions at the SC surface of
the flat mirror located at $z=0$ in Figure 1, we seek solutions of the
vectorized paraxial wave equation \cite{Chen} which result in a longitudinal
electric field vector $\mathbf{E}_{\text{long}}$ satisfying the conducting
boundary conditions at the surface of this mirror, which in turn leads to a 
\emph{longitudinal} component of the electrical force on the charges located
at the surface of the mirror.

The analysis begins with the Maxwell's equations for electromagnetic fields 
\emph{in vacuo} \cite{Jackson} 
\begin{equation}
\nabla \times \mathbf{E}=-\frac{\partial \mathbf{B}}{\partial t}
\end{equation}%
\begin{equation}
\nabla \times \mathbf{B}=+\mu _{0}\varepsilon _{0}\frac{\partial \mathbf{E}}{%
\partial t}
\end{equation}%
\begin{equation}
\nabla \cdot \mathbf{B}=0
\end{equation}%
\begin{equation}
\nabla \cdot \mathbf{E}=0
\end{equation}%
Assuming that all fields have the same complex exponential time dependence $%
\exp \left( -i\omega t\right) $ these equations become%
\begin{equation}
\nabla \times \mathbf{E}=+i\omega \mathbf{B}  \label{faraday}
\end{equation}%
\begin{equation}
\nabla \times \mathbf{B}=-i\omega \mu _{0}\varepsilon _{0}\mathbf{E}
\label{maxwell displacement}
\end{equation}%
\begin{equation}
\nabla \cdot \mathbf{B}=0  \label{div B = 0}
\end{equation}%
\begin{equation}
\nabla \cdot \mathbf{E}=0  \label{div E = 0}
\end{equation}%
Taking the curl of the first Maxwell equation (\ref{faraday}), and using the
second Maxwell equation (\ref{maxwell displacement}), one gets%
\begin{equation}
\nabla \times \left( \nabla \times \mathbf{E}\right) =+i\omega \left( \nabla
\times \mathbf{B}\right) =\omega ^{2}\mu _{0}\varepsilon _{0}\mathbf{E}
\end{equation}%
Using the vector identity%
\begin{equation}
\nabla \times \left( \nabla \times \mathbf{E}\right) =\nabla \left( \nabla
\cdot \mathbf{E}\right) -\nabla ^{2}\mathbf{E}
\end{equation}%
and using the fact that in the vacuum, $\nabla \cdot \mathbf{E}=0$ (i.e.,
that there are no charges present within the volume of the resonator), one
arrives at the Helmholtz equation for the electric field%
\begin{equation}
\nabla ^{2}\mathbf{E}+\omega ^{2}\mu _{0}\varepsilon _{0}\mathbf{E}=\nabla
^{2}\mathbf{E}+k^{2}\mathbf{E}=0  \label{Helmholtz equation for E vector}
\end{equation}%
where%
\begin{equation}
k=\omega \sqrt{\mu _{0}\varepsilon _{0}}=\omega /c
\end{equation}%
is the vacuum wavenumber of the EM wave inside the Fabry-Perot resonator
shown in Figure 1, with $c=1/\sqrt{\mu _{0}\varepsilon _{0}}$ being the
vacuum speed of light.

Likewise, one arrives at the Helmholtz equation for the magnetic field%
\begin{equation}
\nabla ^{2}\mathbf{B}+\omega ^{2}\mu _{0}\varepsilon _{0}\mathbf{B}=\nabla
^{2}\mathbf{B}+k^{2}\mathbf{B}=0  \label{Helmholtz equation for B vector}
\end{equation}

Note that the Fabry-Perot resonator configuration of Figure 1 has a dominant
axis of propagation, namely, the $z$ axis. This suggests that a paraxial
wave approximation for the $z$ component of the electric field might be
useful here.

An important reason for singling out the $z$ component of the electric field
for the Helmholtz equation is that the boundary conditions at the flat SC
mirror require that the longitudinal (or normal)\ component of the electric
field does not vanish at the surface of the mirror, but that the transverse
(or tangential) components of the electric field must vanish at this
conducting (or superconducting) boundary. This boundary condition singles
out a transverse magnetic mode as the fundamental mode of the resonator.

Let us choose the surface of the flat mirror to coincide with the $z=0$
plane. Then from the above conducting boundary conditions, we expect that
the longitudinal component $E_{z}$ is a maximum at $z=0$, and also at $z=-L$
at the conducting surface of the curved mirror, but that the transverse
components of the electric field vanish at these surfaces. We therefore seek
solutions of the $z$ component of the the electric field from Helmholtz
equation (\ref{Helmholtz equation for E vector}), i.e., 
\begin{equation}
\nabla ^{2}E_{z}+k^{2}E_{z}=0  \label{z component Helmholtz equation}
\end{equation}%
that satisfy these boundary conditions. From these solutions for $E_{z}$, we
can derive the transverse components of the electric and magnetic fields of
the modes of the resonator, in a procedure similar to finding the fields of
the transverse magnetic (TM) modes of a microwave resonator \cite{Jackson}.
In other words, the conducting boundary conditions for the configuration of
mirrors shown in Figure 1 demand that the $z$ component of the magnetic
field must be zero everywhere on the surfaces of the mirrors, i.e.,%
\begin{equation}
B_{z}=0
\end{equation}%
a condition that is also required by the Meissner effect in SC's.

It is then natural to try as a solution of the Helmholtz equation (\ref{z
component Helmholtz equation}) the following $Ansatz$ for the $z$ component
of the electric field:%
\begin{equation}
E_{z}\left( x,y,z,t\right) =E_{0}\mathcal{\psi }\left( x,y,z\right) \cos
kz\exp \left( -i\omega t\right) +\text{c.c.}
\label{Ansatz for paraxial solution}
\end{equation}%
where $E_{0}$ is a constant, $\mathcal{\psi }\left( x,y,z\right) $ is a
dimensionless factor with a complex amplitude that varies slowly with $z$
compared to the fast $z$ dependence of the factor $\cos kz$, and ``c.c.," as
usual, is the complex conjugate of the previous term. We have chosen here
the $z$ dependence to be that of a cosine rather than a sine function in
order to satisfy the boundary condition that $E_{z}$ is maximum (i.e., an
anti-node) at $z=0$. This leads to a standing-wave solution for the mode of
the SC resonator.

Therefore neglecting the second derivative of $\mathcal{\psi }\left(
x,y,z\right) $ with respect to $z$, which is assumed to be small compared to
the first derivative with respect to $z$, we arrive at the paraxial wave
equation for the slowly varying amplitude $\mathcal{\psi }$%
\begin{equation}
2ik\frac{\partial \mathcal{\psi }}{\partial z}+\nabla _{t}^{2}\mathcal{\psi }%
=0  \label{paraxial wave equation}
\end{equation}%
where the transverse Laplacian $\nabla _{t}^{2}$ is%
\begin{equation}
\nabla _{t}^{2}=\frac{\partial ^{2}}{\partial x^{2}}+\frac{\partial ^{2}}{%
\partial y^{2}}
\end{equation}

Note that the paraxial wave equation (\ref{paraxial wave equation}) has the
same mathematical form as the time-dependent Schr\"{o}dinger equation%
\begin{equation}
\frac{\hbar }{i}\frac{\partial \mathcal{\psi }}{\partial t}=-\frac{\hbar ^{2}%
}{2m}\nabla ^{2}\mathcal{\psi }  \label{time-dependent S.E.}
\end{equation}%
except that the axial distance $z$ in the paraxial resonator problem has now
been replaced by the time $t$ in the quantum mechanics problem. Since we
know that spreading Gaussian wavepacket solutions are solutions to the
time-dependent Schr\"{o}dinger equation (\ref{time-dependent S.E.}), we
expect that analogous Gaussian solutions will also be solutions to the
paraxial wave equation (\ref{paraxial wave equation}).

For the Fabry-Perot resonator configuration of Figure 1, the procedure is to
first solve the paraxial wave equation (\ref{paraxial wave equation}) for $%
E_{z}$ after setting $B_{z}=0$ everywhere. Then one substitutes these
solutions into the right-hand sides of Maxwell's equations as source terms
to obtain solutions for the transverse fields $E_{x}$ and $E_{y}$.

The lowest-order Gaussian solution to the paraxial wave equation (\ref%
{paraxial wave equation}) can be obtained starting from the following
\textquotedblleft Gaussian\textquotedblright\ $Ansatz$:%
\begin{equation}
\psi \left( x,y,z\right) =\Psi _{G}\left( x,y,z\right) =\exp i\left( P\left(
z\right) +\frac{1}{2}Q\left( z\right) r^{2}\right)  \label{Gaussian Ansatz}
\end{equation}%
where $P\left( z\right) $ and $Q\left( z\right) $ are complex functions of
the axial distance $z$, and where%
\begin{equation}
r^{2}=x^{2}+y^{2}
\end{equation}%
is the square of the radial distance $r$ of a field point displaced away
from the axis of the resonator. Note that $r$ is a real variable.

It follows from this Gaussian $Ansatz$ that%
\begin{equation}
\frac{\partial ^{2}}{\partial x^{2}}\Psi _{G}=iQ\Psi _{G}-Q^{2}x^{2}\Psi _{G}
\end{equation}%
\begin{equation}
\frac{\partial ^{2}}{\partial y^{2}}\Psi _{G}=iQ\Psi _{G}-Q^{2}y^{2}\Psi _{G}
\end{equation}%
so that the paraxial wave equation, which is a PDE, reduces to the ODE%
\begin{equation}
-2k\frac{dP}{dz}-2k\cdot \frac{1}{2}\frac{dQ}{dz}r^{2}+2iQ-Q^{2}r^{2}=0
\end{equation}%
Collecting the coefficients of $r^{2}$ in this equation and setting their
sum equal to zero, one then arrives at the two first-order ODE's%
\begin{eqnarray}
k\frac{dQ}{dz}+Q^{2} &=&0  \label{ODE for Q} \\
-k\frac{dP}{dz}+iQ &=&0  \label{ODE for P}
\end{eqnarray}%
The solution of (\ref{ODE for Q}) can be obtained by integration as follows:%
\begin{equation}
\int \frac{dQ}{Q^{2}}+\frac{1}{k}\int dz=0
\label{first integral of ODE for Q}
\end{equation}%
Transforming variables using%
\begin{equation}
Q=\frac{1}{q}
\end{equation}%
one finds that (\ref{first integral of ODE for Q}) becomes%
\begin{equation}
-\int_{q_{0}}^{q\left( z\right) }dq+\frac{1}{k}\int_{0}^{z}dz=0
\end{equation}%
and therefore that the solution to the ODE for $Q$ (\ref{ODE for Q}) can be
rewritten as follows:%
\begin{equation}
q\left( z\right) -q_{0}=\frac{1}{k}z  \label{q(z)-q_0}
\end{equation}%
It should be kept in mind that since $Q\left( z\right) $ is a complex
function of $z$, so likewise $q\left( z\right) $ will also be a complex
function of $z$.

In analogy with the Gaussian wavepacket-spreading problem in quantum
mechanics, it is natural to impose on the Gaussian $Ansatz$ (\ref{Gaussian
Ansatz}) as an initial condition (i.e., boundary condition) at $z=0$, that
it reduces to the \emph{real} Gaussian function%
\begin{eqnarray}
\Psi _{G}\left( x,y,0\right) =\exp i\left( P\left( 0\right) +\frac{1}{2}%
Q\left( 0\right) r^{2}\right)  \notag \\
=\exp \left( -r^{2}/w_{0}^{2}\right)  \label{initial condition}
\end{eqnarray}%
where the real number $w_{0}$ is the initial Gaussian wavepacket size
evaluated at the $z=0$ plane, which corresponds to the \textquotedblleft
beam waist\textquotedblright\ size at the flat mirror in Figure 1. Then%
\begin{equation}
q_{0}=\frac{1}{Q\left( 0\right) }=-\frac{i}{2}w_{0}^{2}
\end{equation}%
is determined to be an imaginary number. Also, it then follows from the
initial condition (\ref{initial condition}) that the initial value of $%
P\left( z\right) $ must be%
\begin{equation}
P\left( 0\right) =0
\end{equation}%
Therefore the solution for $q\left( z\right) $ (\ref{q(z)-q_0}) becomes%
\begin{equation}
q\left( z\right) =q_{0}+\frac{z}{k}=-\frac{i}{2}w_{0}^{2}+\frac{z}{k}=\frac{%
z-ikw_{0}^{2}/2}{k}=\frac{z-iz_{R}}{k}  \label{solution for q(z)}
\end{equation}%
where the \textquotedblleft Rayleigh range\textquotedblright\ $z_{R}$ is
defined as follows:%
\begin{equation}
z_{R}=kw_{0}^{2}/2  \label{Rayleigh range}
\end{equation}%
The meaning of the Rayleigh range $z_{R}$ is that it is the distance scale
on which the width of the Gaussian beam will have significantly increased
due to diffraction (i.e., wavepacket spreading due to the uncertainty
principle). The paraxial approximation holds when diffraction angle $\theta
_{\text{diffraction}}$\ due to the spreading of the Gaussian beam is
sufficiently small, i.e., when%
\begin{equation}
\theta _{\text{diffraction}}\simeq \frac{\lambda }{w_{0}}<1
\end{equation}%
Therefore the paraxial approximation is satisfied when the Rayleigh range
satifies the condition%
\begin{equation}
z_{R}=kw_{0}^{2}/2=\pi w_{0}^{2}/\lambda \simeq \pi \lambda /\theta _{\text{%
diffraction}}^{2}\geq\lambda/2  \label{condition for paraxial approximation}
\end{equation}

Transforming back from the function $q\left( z\right) $ to the original
function $Q\left( z\right) $, one finds from (\ref{solution for q(z)}) that%
\begin{eqnarray}
Q\left( z\right) &=&\frac{1}{q\left( z\right) }=k\left( \frac{1}{z-iz_{R}}%
\right)=k\left( \frac{z+iz_{R}}{z^{2}+z_{R}^{2}}\right)  \notag \\
&=&k\left( \frac{1}{z+z_{R}^{2}/z}+i\frac{z_{R}}{z^{2}+z_{R}^{2}}\right)
\label{solution for Q(z)}
\end{eqnarray}%
It therefore follows that in the Gaussian $Ansatz$ (\ref{Gaussian Ansatz}),
the exponential factor $\exp \left( \frac{i}{2}Q\left( z\right) r^{2}\right) 
$ can be factorized as follows:%
\begin{eqnarray}
\exp \left( \frac{i}{2}Q\left( z\right) r^{2}\right) &=&\exp \left( \frac{i}{%
2}\frac{kr^{2}}{z+z_{R}^{2}/z}\right)  \notag \\
&\times& \exp \left( -\frac{1}{2}\frac{kr^{2}}{\left(
1+z^{2}/z_{R}^{2}\right) z_{R}}\right)  \label{exp(iQ)} \\
&=&\exp \left( \frac{i}{2}\frac{kr^{2}}{R(z)}\right) \exp \left( -\frac{r^{2}%
}{w^{2}\left( z\right) }\right)  \notag
\end{eqnarray}%
where%
\begin{equation}
R\left( z\right) =z+z_{R}^{2}/z  \label{curvature of phasefront}
\end{equation}%
is the radius of curvature of the phasefront of $\Psi _{G}\left(
x,y,z\right) $ at $z$\thinspace , and where%
\begin{equation}
w^{2}\left( z\right) =w_{0}^{2}\left( 1+\frac{z^{2}}{z_{R}^{2}}\right)
\end{equation}%
is the square of the spreading Gaussian beam width at $z$. Thus one
concludes that the area of the Gaussian beam, as measured by $w^{2}\left(
z\right) $ evaluated at $z=z_{R}$, will double due to diffraction from its
initial value $w_{0}^{2}$ evaluated at the $z=0$ plane.

Next, we shall find the solution of the ODE (\ref{ODE for P}) for $P\left(
z\right) $, starting from the known solution (\ref{solution for Q(z)}) for $%
Q\left( z\right) $, by direct integration, as follows:%
\begin{eqnarray}
P\left( z\right) =\frac{i}{k}\int_{0}^{z}Q\left( z\right) dz=i\int_{0}^{z}%
\frac{1}{z-iz_{R}}dz  \notag \\
=i\ln \left( \frac{z-iz_{R}}{-iz_{R}}\right)  \label{solution for P(z)}
\end{eqnarray}%
Therefore it follows that the first exponential factor in the Gaussian $%
Ansatz$ (\ref{Gaussian Ansatz}) can be expressed as%
\begin{eqnarray}
&&\exp \left( iP\left( z\right) \right)  \notag \\
&=&\exp \left( -\ln \left[ \left( \frac{\sqrt{z^{2}+z_{R}^{2}}}{z_{R}}%
\right) \exp \left( i\varphi _{G}\left( z\right) \right) \right] \right) 
\notag \\
&=&\frac{1}{\sqrt{1+z^{2}/z_{R}^{2}}}\exp \left( -i\arctan \frac{z}{z_{R}}%
\right)
\end{eqnarray}%
where%
\begin{equation}
\varphi _{G}\left( z\right) =-\arctan \frac{z}{z_{R}}  \label{Gouy phase}
\end{equation}%
is called the \textquotedblleft Gouy phase shift\textquotedblright .

Putting everything together, one then finds that the Gaussian $Ansatz$ (\ref%
{Gaussian Ansatz}) becomes%
\begin{eqnarray}
\hspace{-0.1in}\Psi _{G}(x,y,z) &=&\frac{w_{0}}{w\left( z\right) }\exp
\left( -\frac{r^{2}}{w^{2}\left( z\right) }\right)  \notag \\
&&\times \exp \left\{ i\frac{kr^{2}}{2R\left( z\right) }-i\arctan \frac{z}{%
z_{R}}\right\}  \label{Full Gaussian solution}
\end{eqnarray}%
which is a solution of the paraxial wave equation%
\begin{equation}
2ik\frac{\partial \Psi _{G}}{\partial z}+\frac{\partial ^{2}\Psi _{G}}{%
\partial x^{2}}+\frac{\partial ^{2}\Psi _{G}}{\partial y^{2}}=0
\label{paraxial wave equation for Psi_G}
\end{equation}%
and that therefore the full standing-wave solution for the longitudinal
electric field in the transverse-magnetic TM$_{01n}$\ mode of the SC
resonator becomes%
\begin{eqnarray}
E_{z}(x,y,z,t) &=&E_{0}\Psi _{G}\cos kz\exp \left( -i\omega t\right) +\text{%
c.c.}  \notag \\
&=&E_{0}\frac{w_{0}}{w\left( z\right) }\exp \left( -\frac{r^{2}}{w^{2}\left(
z\right) }\right)  \notag \\
&&\times \exp \left\{ i\frac{kr^{2}}{2R\left( z\right) }-i\arctan \frac{z}{%
z_{R}}\right\}  \notag \\
&&\times \cos kz\exp \left( -i\omega t\right) +\text{c.c.}
\label{full solution for E_z(x,y,z,t)}
\end{eqnarray}%
The first boundary condition at the flat mirror at $z=0$ will be satisfied,
since%
\begin{equation}
\cos kz=1\text{ at }z=0
\end{equation}%
and therefore at $z=0$ 
\begin{equation}
E_{z}(x,y,0,t)=E_{0}\exp \left( -\frac{r^{2}}{w_{0}^{2}}\right) \exp \left(
-i\omega t\right) +\text{c.c.}
\end{equation}%
i.e., that the longitudinal electric field has a standing-wave anti-node at $%
z=0$. If one further assumes that the curvature of the curved mirror in
Figure 1 matches the curvature of the curved phase front $R\left( z\right) $
given by (\ref{curvature of phasefront}) evaluated at $z=-L$, then the
second boundary condition at the curved mirror at $z=-L$ will be satisfied
when 
\begin{equation}
\cos kL=\pm 1\text{ for }kL=n\pi \text{ where }n=1,2,3,...  \label{cos kL}
\end{equation}%
i.e., that the longitudinal electric field has a standing-wave anti-node at $%
z=-L$ (we neglect here a small correction factor arising from the Gouy phase
shift). Therefore the standing-wave anti-node condition (\ref{cos kL})
determines the TM$_{01n}$ eigenmode frequencies of the SC resonator depicted
in Figure 1 through the relationship%
\begin{equation}
\omega _{n}=ck_{n}=2\pi \cdot n\frac{c}{2L}\text{ where }n=1,2,3,...
\end{equation}
Hence the fundamental eigenmode, i.e., the one with the lowest resonance frequency, corresponds to the mode number $n=1$, i.e., the TM$_{011}$ mode.

In order to find the transverse components $E_{x}(x,y,z,t)$ and $%
E_{y}(x,y,z,t)$, let us return to Maxwell's equations. Since no free charges
are present in the vacuum between the two SC mirrors of Figure 1, it follows
that $\nabla \cdot \mathbf{E}=0$, and therefore that%
\begin{equation}
\frac{\partial E_{x}}{\partial x}+\frac{\partial E_{y}}{\partial y}=-\frac{%
\partial E_{z}}{\partial z}  \label{div E=0}
\end{equation}%
From the paraxial approximation condition (\ref{condition for paraxial
approximation}), one concludes that the leading term on the right-hand side
will be given by%
\begin{equation}
-\frac{\partial E_{z}}{\partial z}\doteq E_{0}\Psi _{G}(x,y,z)k\sin kz\exp
\left( -i\omega t\right) +\text{c.c.}
\end{equation}%
This suggests that we try as trial solutions%
\begin{equation}
E_{x}=x\cdot \frac{E_{0}}{2}\Psi _{G}k\sin kz\exp \left( -i\omega t\right) +%
\text{c.c.}  \label{trial solution for E_x}
\end{equation}%
\begin{equation}
E_{y}=y\cdot \frac{E_{0}}{2}\Psi _{G}k\sin kz\exp \left( -i\omega t\right) +%
\text{c.c}  \label{trial solution for E_y}
\end{equation}%
because then%
\begin{equation}
\frac{\partial \left( x\Psi _{G}\right) }{\partial x}=\Psi _{G}+x\frac{%
\partial \Psi _{G}}{\partial x}=\Psi _{G}+x\frac{i}{2}Q\cdot x\Psi _{G}
\end{equation}%
where the last term can be neglected in the paraxial approximation, so that
the leading terms are%
\begin{equation}
\frac{\partial \left( x\Psi _{G}\right) }{\partial x}\doteq \Psi _{G}
\end{equation}%
\begin{equation}
\frac{\partial \left( y\Psi _{G}\right) }{\partial y}\doteq \Psi _{G}
\end{equation}%
Therefore (\ref{div E=0}) becomes%
\begin{equation}
\frac{1}{2}k\frac{\partial \left( x\Psi _{G}\right) }{\partial x}+\frac{1}{2}%
k\frac{\partial \left( y\Psi _{G}\right) }{\partial y}\doteq k\Psi _{G}
\end{equation}%
which holds true in the paraxial approximation.

As a further check of the validity of the above trial solutions (\ref{trial
solution for E_x}) and (\ref{trial solution for E_y}), let us verify that
Faraday's law in component form%
\begin{equation}
\frac{\partial E_{x}}{\partial y}-\frac{\partial E_{y}}{\partial x}=-\frac{%
\partial B_{z}}{\partial t}=0  \label{faraday z component}
\end{equation}%
is satisfied. One finds that%
\begin{equation}
\frac{\partial E_{x}}{\partial y}\propto \frac{\partial \left( x\Psi
_{G}\right) }{\partial y}=x\cdot 2y\cdot iQ\Psi _{G}
\end{equation}%
\begin{equation}
\frac{\partial E_{y}}{\partial x}\propto \frac{\partial \left( y\Psi
_{G}\right) }{\partial x}=y\cdot 2x\cdot iQ\Psi _{G}
\end{equation}%
cancel each other, so that we see that (\ref{faraday z component}) is indeed
satisfied. We therefore conclude that, in the paraxial approximation, (\ref%
{trial solution for E_x}) and (\ref{trial solution for E_y}) are indeed the
unique solutions for the transverse components $E_{x}$ and $E_{y}$ that
correspond to the paraxial TM$_{01n}$ Gaussian-beam solution $E_{z}$ given
by (\ref{full solution for E_z(x,y,z,t)}).

The two transverse magnetic field components $B_{x}$ and $B_{y}$ can then be
gotten from the two Maxwell equations%
\begin{equation}
\frac{\partial B_{x}}{\partial x}+\frac{\partial B_{y}}{\partial y}=-\frac{%
\partial B_{z}}{\partial z}=0  \label{div B = 0 in component form}
\end{equation}%
\begin{equation}
\frac{\partial B_{x}}{\partial y}-\frac{\partial B_{y}}{\partial x}=\mu
_{0}\varepsilon _{0}\frac{\partial E_{z}}{\partial t}=-i\omega \mu
_{0}\varepsilon _{0}E_{z}  \label{Ampere-Maxwell equation in component form}
\end{equation}%
By inspection, the solutions for $B_{x}$ and $B_{y}$ of (\ref{div B = 0 in
component form}) and (\ref{Ampere-Maxwell equation in component form}) are%
\begin{equation}
B_{x}=y\cdot \frac{E_{0}}{2c}\Psi _{G}k\cos kz\left( -ie^{-i\omega t}\right)
+\text{c.c.}
\end{equation}%
\begin{equation}
B_{y}=-x\cdot \frac{E_{0}}{2c}\Psi _{G}k\cos kz\left( -ie^{-i\omega
t}\right) +\text{c.c.}
\end{equation}

\begin{figure}[tbh]
\centering
\includegraphics[angle=0,width=.4\textwidth]{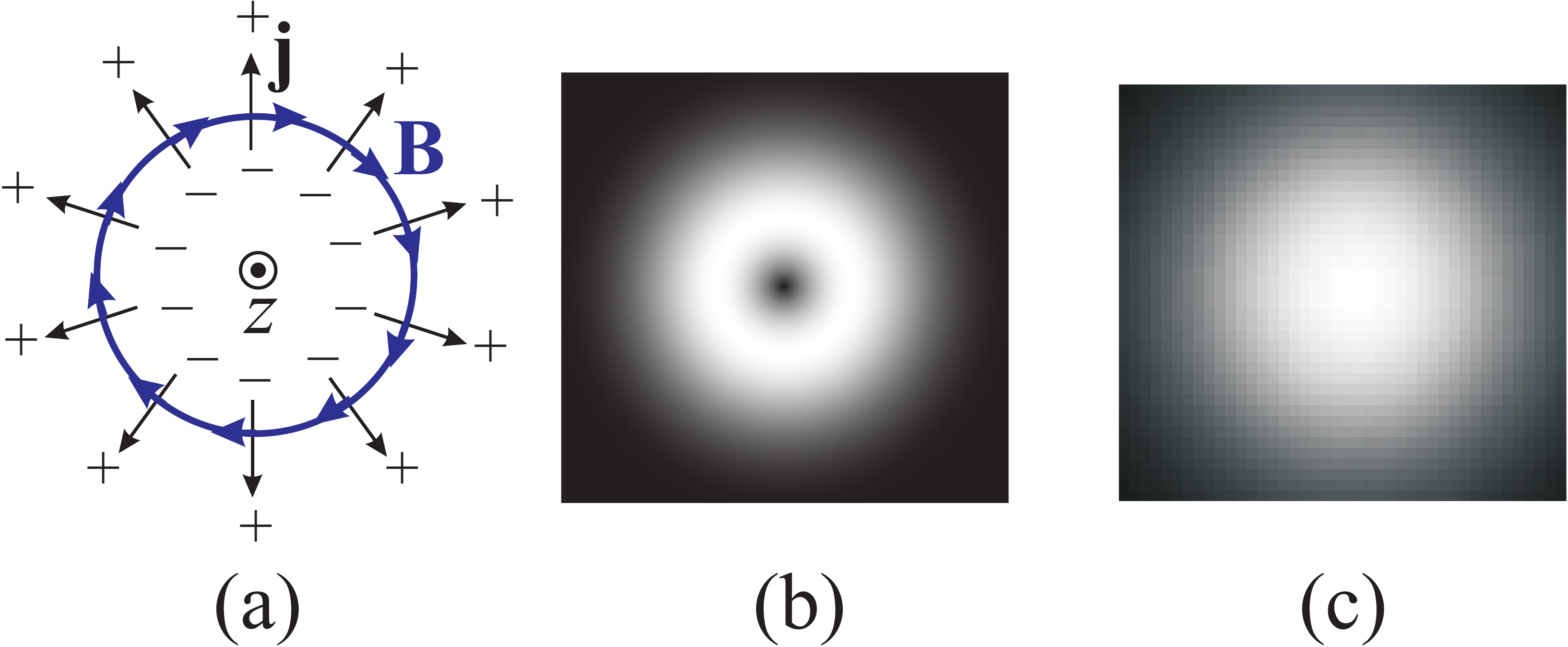}
\caption{(a) Sketch of a snapshot of the pattern of supercurrents $\mathbf{j}
$ (black arrows) flowing on the surface of the SC flat mirror at $z=0$ of
Figure 1, and their associated magnetic field vectors $\mathbf{B}$ (blue
arrows) immediately outside of the SC surface. This pattern of currents and
fields is associated with the circularly symmetric TM$_{01n}$ mode solution (%
\protect\ref{full solution for E_z(x,y,z,t)}). (b) Transverse intensity
profile of this mode (i.e., the sum of the absolute squares of $E_{x}$ and $%
E_{y}$ given by (\protect\ref{trial solution for E_x}) and (\protect\ref%
{trial solution for E_y}), respectively). (c) Longitudinal intensity profile
(i.e., the absolute square of $E_{z}$ given by (\protect\ref{full solution
for E_z(x,y,z,t)})).}
\label{fig:TM01-mode}
\end{figure}


The above solutions can be depicted as the circularly symmetric, transverse
magnetic TM$_{01n}$ mode pattern sketched in Figure 6(a), which is a
snapshot taken looking down the $z$ axis of the hemiconfocal resonator at
the $z=0$ plane of the flat SC mirror of Figure 1, at the moment of maximum
magnetic field. In Figure 6(b), the transverse intensity pattern of this
mode is depicted. The circular symmetry of this pattern can be readily
understood by taking the absolute square of the transverse electric fields
of this mode, which are given by (\ref{trial solution for E_x}) and (\ref%
{trial solution for E_y}), i.e.,%
\begin{equation}
E_{x}\propto x=r\cos \theta
\end{equation}%
\begin{equation}
E_{y}\propto y=r\sin \theta
\end{equation}%
so that%
\begin{equation}
\left\vert E_{x}\right\vert ^{2}+\left\vert E_{y}\right\vert ^{2}\propto
r^{2}
\end{equation}%
which is clearly independent of the azimuthal angle $\theta $. Figure 6(c)
shows the longitudinal intensity profile of $\vert E_z \vert ^2$.

\section{Appendix B: Simple harmonic oscillator model for the microwave
cavity-pellicle mirror system}

Here it is shown that the superconducting (SC) microwave cavity with a
pellicle end-mirror (see Figures 1 and 4) can be modeled as a simple
harmonic oscillator whose loaded quality factor $Q_{\text{loaded}}$ is
approximately given by the quality factor of the SC cavity $Q_{s}$.

Let the pellicle end-mirror consist of a thin SC film deposited on a thin,
light, flexible diaphragm, which is sufficiently thin so that it can easily
be driven into mechanical motion. Furthermore, suppose that the SC film is
electrostatically charged with a net DC charge $q$. We assume that the
charge $q$ which resides on the surface of the film is so tightly bound (via
the Coulomb force) to the metallic film that when the charge $q$ moves, the
film will co-move with it \cite{Raynote0}. Then the longitudinal electric
field $\mathbf{E}_{z}$ at the surface of the SC film will lead to the
instantaneous force 
\begin{equation}
\mathbf{F}_{z}\left( t\right) =q\mathbf{E}_{z}(t)+\mathbf{F}_{\text{rad}%
}\left( t\right)  \label{eq:FT}
\end{equation}%
where $\mathbf{E}_{z}$ is the longitudinal electric field at the surface of
the SC film, and where the force on the film due to radiation pressure is
given by%
\begin{equation}
\mathbf{F}_{\text{rad}}\left( t\right) =\frac{1}{2}\varepsilon _{0}\mathbf{E}%
_{z}(t)^{2}\mathcal{A}\propto \mathbf{E}_{z}(t)^{2}  \label{F_rad(t)}
\end{equation}%
where $\epsilon _{0}$ is the permittivity of free space, and $\mathcal{A}$
is the area of the film over which $\mathbf{E}_{z}(t)$ is nonvanishing,
i.e., the Gaussian-beam-waist area of Figure 6(c). Since the radiation force 
$\mathbf{F}_{\text{rad}}\left( t\right) $ scales quadratically with the
electric field at the surface of the film, while the Coulomb force $q\mathbf{%
E}_{z}(t)$ scales linearly, there exists a maximum electric field strength $%
E_{\text{max}}$ such that if $\left\vert E_{z}\left( t\right) \right\vert
<E_{\text{max}}$, then the Coulomb force $q\mathbf{E}_{z}(t)$ dominates over
the radiation force $\mathbf{F}_{\text{rad}}\left( t\right) $ . Comparing (%
\ref{eq:FT}) and (\ref{F_rad(t)}), one finds that 
\begin{equation*}
E_{\text{max}}=\frac{2q}{\epsilon _{0}\mathcal{A}}
\end{equation*}%
Expressing this condition in terms of a maximum externally applied
\textquotedblleft seed\textquotedblright\ power ($\mathcal{P}_{\text{ext}}^{%
\text{max}}$) and the maximum circulating power in the cavity ($\mathcal{P}_{%
\text{cav}}^{\text{max}}$) we find 
\begin{align}
& \mathcal{P}_{\text{ext}}^{\text{max}}=\frac{U_{0}\omega }{4Q_{s}}\approx 7~%
\text{nW} \\
& \mathcal{P}_{\text{cav}}^{\text{max}}=\frac{q^{2}c}{\epsilon _{0}\mathcal{A%
}}\approx 43~\text{W}
\end{align}%
where we have assumed perfect coupling between the externally applied
\textquotedblleft seed\textquotedblright\ power and the cavity, $\mathcal{A}%
=\pi w_{0}^{2}$ is the cross-sectional area given by the beam waist (with $%
w_{0}=1$ cm) of the TM$_{011}$ mode in cavity, $q\approx 20$ pC for 100
Volts DC, $\omega =2\pi \times 12$ GHz, and $Q_{s}=10^{10}$ \cite%
{kuhr07}. For the rest of this analysis, we assume that we are in the regime
where the circulating power in the cavity is sufficiently less than 43 W,
or, more generally, that it is less than $\mathcal{P}_{\text{cav}}^{\text{max%
}}$, so that the radiation force is negligible, and (\ref{eq:FT}) becomes 
\begin{equation}
\mathbf{F}_{z}\left( t\right) \approx q\mathbf{E}_{z}(t)
\end{equation}

At high (i.e., microwave) frequencies we take the approximation that the
pellicle end-mirror behaves like a free mass, so that the equation of motion
for the pellicle mirror is given as 
\begin{equation}
m\frac{d^{2}x}{dt^{2}}=qE(t),
\end{equation}%
where we drop the subscript $z$ from $\mathbf{E}_{z}(t)$ for convenience,
and we switch from $z$ to the variable $x$ to denote the displacement of the
oscillating mass $m$ from equilibrium. The time-dependent part of the
longitudinal electric field at the surface of the SC film can be described
as a harmonically time-varying field given by 
\begin{equation*}
E(t)=\mathcal{E}(t)e^{-i\omega t}+\text{c.c.}
\end{equation*}%
where $\mathcal{E}(t)$ is a slow varying amplitude. It follows that the
displacement of the charged mirror is given by 
\begin{equation}
x(t)=-\frac{q}{m\omega ^{2}}E(t).  \label{eq:Dis}
\end{equation}%
Observe that the displacement $x(t)$ is \textit{linear} with the
longitudinal electric field $E(t)$ evaluated at the surface of the flat
mirror in the SC resonator. Now suppose that the SC resonator is in steady
state and filled with some constant input power from some \textquotedblleft
seed" microwaves so that the pellicle end-mirror displacement is given by (%
\ref{eq:Dis}). If we then shut off the injected \textquotedblleft
seed\textquotedblright\ power, the SC resonator's electric field will decay
exponentially with time. Hence, the displacement of the pellicle end-mirror
will also decay exponentially with time. It now suffices to show that the
electric field $E(t)$ in the resonator can be described by a simple harmonic
oscillator. The equation of motion for the undriven simple harmonic
oscillator is 
\begin{equation}
\frac{d^{2}x}{dt^{2}}+2\gamma \frac{dx}{dt}+\omega ^{2}x=0
\end{equation}%
where $\gamma $ is the decay parameter of the oscillator. Using (\ref{eq:Dis}%
) we arrive at an equivalent simple harmonic motion equation for the field
in the cavity 
\begin{equation}  \label{eq:ESHO}
\frac{d^{2}E}{dt^{2}}+2\gamma \frac{dE}{dt}+\omega ^{2}E=0
\end{equation}%
where $E$ is the electric field evaluated at the surface of the moving
mirror, where we interpret $2\gamma $ as the FWHM of the SC cavity
resonance, and where $\omega $ is the resonance frequency of the SC
resonator. (Note that (\ref{eq:ESHO}) also follows from the Helmholtz
analysis for a lossy resonator.) In the slowly varying amplitude
approximation, (\ref{eq:ESHO}) reduces to a first order linear differential
equation for the slowly varying amplitude
\begin{equation}
\frac{d\mathcal{E}}{dt}+\gamma\mathcal{E} =0  \label{eq:LODE}
\end{equation}%
whose solution is 
\begin{equation}
\mathcal{E}(t)=E_{0}e^{-{\gamma}t}=E_{0}e^{-\omega t/2Q_{s}}  \label{eq:Sol}
\end{equation}%
where $E_{0}$ is the initial electric field amplitude at the surface of the
mirror, and $Q_{s}=\omega /2\gamma $ is the SC resonator's intrinsic quality
factor. Therefore, the field in the SC microwave resonator decays like a
simple harmonic oscillator with a time constant that is proportional to the
quality factor of the cavity. Furthermore, since the displacement of the
pellicle end-mirror is linear with the field inside the resonator, it must
also decay like a simple harmonic oscillator. Note that with (\ref{eq:Sol})
we can write the field inside the resonator as 
\begin{equation}
E(t)=E_{0}e^{-\omega t/2Q_{s}}e^{-i\omega t}+\text{c.c.}
\end{equation}%
which is the well-known exponentially decaying solution with the ringdown
time 
\begin{equation*}
\tau _{r}=2Q_{s}/\omega
\end{equation*}%
of the resonator.

Finally we take into consideration the effect that driving the pellicle
end-mirror with \textquotedblleft seed\textquotedblright\ radiation has on
the quality factor $Q_{s}$ of the SC resonator. The loaded quality factor $%
Q_{\text{loaded}}$, where the loading refers to the power loss due to the
simple harmonic motion of the charged mirror, is given by 
\begin{equation}
Q_{\text{loaded}}=\frac{\omega U_{0}}{\mathcal{P}_{\text{loss}}}
\label{eq:Q}
\end{equation}%
where $U_{0}$ is the energy stored in the cavity, and $\mathcal{P}_{\text{%
loss}}$ is the total power loss in the cavity given by%
\begin{equation}
\mathcal{P}_{\text{loss}}=\mathcal{P}_{c}+\mathcal{P}_{\text{mirror}}
\end{equation}%
where $\mathcal{P}_{c}$ is the intrinsic power loss of the SC resonator and
is related to the resonator's intrinsic quality factor by $Q_{s}=\omega
U_{0}/\mathcal{P}_{c}$, and where $\mathcal{P}_{\text{mirror}}$ is the
average power loss due to the motion of the charged pellicle end-mirror. From (\ref%
{eq:Q}) one finds that 
\begin{equation}
Q_{\text{loaded}}=\frac{Q_{s}Q_{\text{SHM}}}{Q_{s}+Q_{\text{SHM}}}
\label{eq:EQ}
\end{equation}%
where $Q_{\text{SHM}}\equiv \omega U_{0}/P_{\text{mirror}}$ is the
contribution to the quality factor arising from simple harmonic motion
(SHM). Although it is possible that some or all of the power loss that goes
into the simple harmonic motion of the charged mirror is converted into
electromagnetic radiation power which goes back into the SC resonator, it is
instructive to account for it. The average power loss due to the moving
pellicle end-mirror is 
\begin{equation}
\mathcal{P}_{\text{mirror}}=\left\langle \mathbf{F}\cdot \mathbf{v}%
\right\rangle =\frac{q^{2}E_{0}^{2}}{2m\omega }.
\end{equation}%
The electric field is calculated by assuming some externally applied
\textquotedblleft seed\textquotedblright\ power $\mathcal{P}_{\text{ext}}$
is injected into the SC resonator. In steady state, the energy in the cavity $%
U_{0}$ is 
\begin{equation}
U_{0}=\frac{4\beta \mathcal{P}_{\text{ext}}}{(1+\beta )^{2}}\frac{Q_{s}}{%
\omega }
\end{equation}%
where $\beta $ is a coupling parameter of the input/output hole in Figure 1
and is assumed to be unity, $\beta \equiv 1$. It follows that the amplitude
of the electric field inside the cavity $E_{0}$ is given by 
\begin{equation}
E_{0}^{2}=\frac{8\mathcal{P}_{\text{ext}}Q_{s}}{m\omega \mathcal{V}\epsilon
_{0}}
\end{equation}%
where $\mathcal{V}$ is the effective Gaussian-beam volume of the SC
resonator. Hence, the average power loss from the pellicle end-mirror is 
\begin{equation}
\mathcal{P}_{\text{mirror}}=\frac{4q^{2}\mathcal{P}_{\text{ext}}Q_{s}}{%
m\omega ^{2}\mathcal{V}\epsilon _{0}}.
\end{equation}

Assuming an external applied \textquotedblleft seed\textquotedblright\ power
of $\mathcal{P}_{\text{ext}}=500$ pW, $q=20$ pC, $Q_{s}=10^{10}$, $%
m=2$ mg, $\omega =2\pi \times 12$ GHz, and an effective Gaussian-beam volume
of $\mathcal{V}=1.25$ cm$^{3}$, one finds 
\begin{equation}
\mathcal{P}_{\text{mirror}}\approx 6\times 10^{-20}~\text{W}
\end{equation}

This power yields 
\begin{equation}
Q_{\text{SHM}}\approx 4\times 10^{21}.
\end{equation}%
Since $Q_{s}\ll Q_{\text{SHM}}$, it follows from (\ref{eq:EQ}) that to an
extremely good approximation 
\begin{equation*}
Q_{\text{loaded}}\approx Q_{s}
\end{equation*}

\section{Appendix C: Simple Solution to a Fabry-Perot with a Harmonically
Moving End-mirror in Quasi Steady State}

We consider a Fabry-Perot under quasi steady state conditions and ignore
transients during the build up of the modes as those discussed in \cite%
{lawrence99}. In a quasi steady state, the harmonically moving end-mirror
generates two Doppler side bands, \textquotedblleft
Stokes\textquotedblright\ and \textquotedblleft
anti-Stokes\textquotedblright\ \cite{cooper80}. The \textquotedblleft
anti-Stokes\textquotedblright\ sideband is suppressed via the
\textquotedblleft Double" Fabry-Perot scheme as illustrated in Figures 3 and
5.

The radiation force is given by 
\begin{equation}
F(t)=\frac{1}{2}\epsilon _{0}|E|^{2}\mathcal{A}  \label{eq:radforce}
\end{equation}%
where $\epsilon _{0}$ is the permitivity of free space, $\mathcal{A}$ is the
cross sectional area of the Gaussian beam, and $E$ is the total electric
field in the Fabry-Perot. Modeling the moving end-mirror as a simple
harmonic oscillator as depicted in Figure 2, we have 
\begin{equation}
\ddot{x}+2\gamma _{s}\dot{x}+\omega _{s}^{2}x=\frac{F(t)}{m},  \label{eq:sho}
\end{equation}%
where $x$ is the displacement of the simple harmonic oscillator from
equilibrium, $\omega _{s}$ is the natural oscillator frequency, $m$ is its
mass, and $2\gamma _{s}$ is the FWHM. Using the slowly varying amplitude
approximation in quasi steady state 
\begin{equation}
x=X(t)e^{-i\omega _{s}t}+\text{c.c.}
\end{equation}%
the left side of (\ref{eq:sho}) becomes 
\begin{equation}
-2i\omega _{s}\left( \frac{dX}{dt}+\gamma _{s}X\right) e^{-i\omega _{s}t}+%
\text{c.c.}
\end{equation}%
where we have assumed that $2\left( \gamma _{s}-i\omega _{s}\right) \approx
-2i\omega _{s}$. The right hand side can be expanded in terms of the fields%
\begin{equation}
E_{i}=\mathcal{E}_{0i}(t)e^{-i\omega _{i}t}+\text{c.c., ~}E_{p}=\mathcal{E}_{0p}e^{-i\omega
_{p}t}+\text{c.c.}
\end{equation}%
where the former is the \textquotedblleft Stokes\textquotedblright\ or
\textquotedblleft idler\textquotedblright\ term and the latter is the
\textquotedblleft pump\textquotedblright\ mode, $\mathcal{E}_{0i}(t)$ being a slowly
varying amplitude in quasi steady state, but $\mathcal{E}_{0p}$ being a constant, in
the \textquotedblleft undepleted pump\textquotedblright\ approximation.
Taking the beat terms and neglecting the nonresonant terms $|E_{i}|^{2}$ and 
$|E_{p}|^{2}$ we have 
\begin{align}
F(t)=& \epsilon _{0}\mathcal{A}[\mathcal{E}_{0i}^{\ast }\mathcal{E}_{0p}e^{-i(\omega _{p}-\omega
_{i})t}  \notag \\
& +\mathcal{E}_{0i}\mathcal{E}_{0p}e^{-i(\omega _{p}+\omega _{i})t}+\text{c.c.}]
\end{align}%
Equating both sides and multiplying by $e^{i\omega _{s}t}$ we find 
\begin{eqnarray}
&&-2i\omega _{s}\left( \frac{dX}{dt}+\gamma _{s}X\right) +\text{c.c.}%
(\propto e^{2i\omega _{s}t})  \notag \\
&=&\frac{\epsilon _{0}\mathcal{A}}{m}[\mathcal{E}_{0i}^{\ast }\mathcal{E}_{0p}e^{-i(\omega
_{p}-\omega _{i}-\omega _{s})t}+\mathcal{E}_{0i}\mathcal{E}_{0p}e^{2i\omega _{i}t}  \notag \\
&&+\text{c.c.}(\propto e^{i2\omega _{s}t},e^{i2\omega _{p}t})],
\end{eqnarray}%
where $\omega _{p}-\omega _{i}\approx \omega _{s}$. Here we employ the fact
that this is a linear system and observe that only the force at resonance
will be the main driving force of the mechanical oscillator; hence, we can
neglect off-resonance terms (in the rotating-wave approximation), and write the driven oscillator equation as 
\begin{equation}
\frac{dX}{dt}+\gamma _{s}X=\frac{i\epsilon _{0}\mathcal{A}}{2\omega _{s}m}%
\mathcal{E}_{0i}^{\ast }\mathcal{E}_{0p}e^{-i(\omega _{p}-\omega _{i}-\omega _{s})t}
\label{eq:meom}
\end{equation}%
Following \cite{braginsky01}, let us define the fields as 
\begin{align}
E_{i}\left( t\right) & =A_{i}(D_{i}\left( t\right) e^{-i\omega _{i}t}+\text{%
c.c.}) \\
E_{p}& =A_{p}(D_{p}e^{-i\omega _{p}t}+\text{c.c.})
\end{align}%
where $A_{i},~A_{p}$ are normalized so that the total energy stored in each
mode is $U_{p,i}=2\omega _{p,i}|D_{p,i}|^{2}$, and $D_{i}\left( t\right) $
is a slowly varying complex amplitude, but $D_{p}$ is a constant, in the
\textquotedblleft undepleted pump\textquotedblright\ approximation.
Calculating the energies in each mode we find 
\begin{equation}
U_{p,i}=\frac{\mathcal{V}}{2}\epsilon _{0}\left\langle E^{2}\right\rangle
=A_{p,i}^{2}\epsilon _{0}|D_{p,i}|^{2}\mathcal{A}L=2\omega
_{p,i}^{2}|D_{p,i}|^{2}
\end{equation}%
where $\mathcal{V}=\mathcal{A}L$ is the volume of the cavity. Solving for $%
A_{p,i}$ 
\begin{equation}
A_{p,i}=\omega _{p,i}\sqrt{\frac{2}{\epsilon _{0}\mathcal{V}}}
\end{equation}%
Inserting the normalization%
\begin{equation}
\mathcal{E}_{0i}^{\ast }=A_{i}D_{i}^{\ast }\text{ and}~\mathcal{E}_{0p}=A_{p}D_{p}
\end{equation}%
into equation \eqref{eq:meom} we arrive at the result \cite{braginsky01},
eq. (2), 
\begin{equation}
\frac{dX}{dt}+\gamma _{s}X=\frac{i\omega _{p}\omega _{i}}{%
m\omega _{s}L}D_{p}D_{i}^{\ast }e^{-i\Delta \omega t}  \label{eq:beomx}
\end{equation}%
where $\Delta \omega \equiv \omega _{p}-\omega _{i}-\omega _{s}$.

To arrive at the equation for the \textquotedblleft
Stokes\textquotedblright\ or \textquotedblleft idler\textquotedblright\
field in the cavity we note that the \textquotedblleft
Stokes\textquotedblright\ mode is generated from the main \textquotedblleft
pump\textquotedblright\ mode so that the \textquotedblleft
Stokes\textquotedblright\ field is proportional to the oscillator amplitude.
Recall that we are assuming quasi steady state conditions. The electric
field reflected from a moving mirror is given by \cite{cooper80} 
\begin{equation}
E_{p}=\mathcal{E}_{0p}e^{-i\omega _{p}t}e^{2ik_{p}x}+\text{c.c.}
\end{equation}%
For small $x$, we find 
\begin{equation}
E_{p}=\mathcal{E}_{0p}e^{-i\omega _{p}t}(1+2ik_{p}x)+\text{c.c.}
\end{equation}%
where the last term leads to the Doppler-generated electric field 
\begin{equation}
E_{\text{Doppler}}\equiv 2ik_{p}x\mathcal{E}_{0p}e^{-i\omega _{p}t}+\text{c.c.}
\end{equation}%
Since we assume quasi steady state conditions, the \textquotedblleft
pump\textquotedblright\ optical mode is the main source of the
\textquotedblleft Stokes\textquotedblright\ optical mode. As is well known,
the Fabry-Perot cavity field obeys a recursion relation; for example see 
\cite{rakhmanov02,lawrence99}. Hence 
\begin{equation}
E_{i}(t+\tau )=E_{\text{Doppler}}(t+\tau )+\mathcal{R}P_{f}E_{i}(t)
\end{equation}%
where $\tau =2L/c$ is the round trip time, $\mathcal{R}$ is the power
reflectivity (i.e., the absolute square of the reflection coefficient) of
the end mirrors, $P_{f}$ is a propagation factor \cite%
{rakhmanov02,lawrence99} which accounts for the phase accumulated by the
beam one round trip earlier, and $E_{\text{Doppler}}$ is the Doppler
electric field generated from the \textquotedblleft pump\textquotedblright\
mode. Putting in the slowly varying displacements and fields as before%
\begin{eqnarray}
x &=&X\left( t\right) e^{-i\omega _{s}t}+\text{c.c.} \\
E_{i} &=&\mathcal{E}_{0i}\left( t\right) e^{-i\omega _{i}t}+\text{c.c.} \\
\text{ }E_{p} &=&\mathcal{E}_{0p}e^{-i\omega _{p}t}+\text{c.c.}
\end{eqnarray}%
and taking the resonant terms, we find 
\begin{equation}
\frac{d\mathcal{E}_{0i}}{dt}+\gamma _{i}\mathcal{E}_{0i}=\frac{i}{\tau }%
X^{\ast }\mathcal{E}_{0p}e^{-i\Delta \omega t}  \label{eq:meomd2}
\end{equation}%
where we make use of $\Delta \omega \equiv \omega _{p}-\omega _{i}-\omega
_{s}$. Inserting the normalization used in \cite{braginsky01} we find 
\begin{equation}
\frac{dD_{i}}{dt}+\gamma _{i}D_{i}=\frac{i}{L}\frac{\omega
_{p}^{2}}{\omega _{i}}X^{\ast }D_{p}e^{-i\Delta \omega t}
\end{equation}%
where we have used the fact that $\tau =2L/c$, and $k_{p}c=\omega _{p}$.
With the approximation that 
\begin{equation}
\frac{\omega _{p}^{2}}{\omega _{i}}=\frac{\omega _{p}^{2}}{\omega
_{p}-\omega _{s}}\approx \frac{\omega _{p}^{2}}{\omega _{p}}=\omega _{p}
\end{equation}%
(i.e. $\omega _{p}\gg \omega _{s}$) we find, 
\begin{equation}
\frac{dD_{i}}{dt}+\gamma _{i}D_{i}=\frac{iX^{\ast }D_{p}\omega _{p}}{L}%
e^{-i\Delta \omega t}.  \label{eq:beomd2}
\end{equation}%
This result is consistent with \cite{braginsky01}, eq. (1). For the detailed
solutions to the coupled differential equations \eqref{eq:beomx} and %
\eqref{eq:beomd2} and the condition for parametric amplification the reader
is referred to \cite{braginsky01}. The threshold condition is 
\begin{equation}
\frac{2U_{p}Q_{i}Q_{s}}{mL^{2}\omega _{s}^{2}}>1
\end{equation}%
where $U_{p}$ is the energy stored in the \textquotedblleft
pump\textquotedblright\ mode, $L$ is the length of the Fabry-Perot, $m$ is
the effective mass, and the $Q$'s are the quality factors.

Note that the $Q$'s are defined as follows 
\begin{align}
& Q_{s}=\frac{\omega _{s}}{2\gamma _{s}},~\text{\textquotedblleft
Mechanical\textquotedblright\ Oscillator} \\
& Q_{i}=\frac{\omega _{i}}{2\gamma _{i}},~\text{\textquotedblleft
Stokes\textquotedblright\ mode in FP} \\
& Q_{p}=\frac{\omega _{p}}{2\gamma _{p}},~\text{\textquotedblleft
Pump\textquotedblright\ mode in FP}
\end{align}%
where the $\gamma $'s correspond to the relaxation rate for each of the
modes (i.e HWHM). 


\begin{thebibliography}{99}
\bibitem{nonadiabatic} Due to the finite time delay between successive
bounces of the waves between the fixed mirror and the moving mirror, the
action of the moving mirror upon the radiation contained in the cavity is
non-adiabatic.
\bibitem{luisnote0} A reflection coefficient of $0.9999$ can be achieve by
using 9 double layer stack of ZnS ($n=2.35$) and MgF$_{2}$ ($n=1.38$)
coatings on mylar ($n=1.6$). For 700nm light this yields a thickness of 3.15
microns.
\bibitem{luisnote1} We select $Q_s$ as a figure of merit based on published
experimental work for superconducting cavities. However, $Q_s$ depends
sensitively on the fabrication procedure, and the design of the microwave
cavity.
\bibitem{luisnote2} The mass is calculated from a 1.5 micron Nb coating,
3.15 micron multilayer dielectric reflection coating, and a 2 micron mylar
drum thickness. As an order of magnitude we assume that in this high
frequency regime the effective mass is on the order of the free mass (i.e.
the pellicle acts as a free mass).
\bibitem{Raynote0} This is justified because the binding energy of the net
charge of electrons to the surface of the metal is on the order of tens of
eV, whereas the simple harmonic motion of the mirror corresponds to an
energy on the order of meV.
\end{thebibliography}
\end{document}